\documentclass[11pt,a4paper]{article}
\usepackage{jheppub}
\usepackage{graphicx, bm, color, amssymb, amsmath,slashed,xcolor}
\usepackage{epstopdf}
\usepackage{multirow}
\usepackage{float}
\usepackage[caption=false]{subfig}
\usepackage{tabularx}
\usepackage{hyperref}
\usepackage[normalem]{ulem}
\newcommand{\lsim}{\mbox{\raisebox{-.6ex}{~$\stackrel{<}{\sim}$~}}}
\newcommand{\gsim}{\mbox{\raisebox{-.6ex}{~$\stackrel{>}{\sim}$~}}}

\newcolumntype{L}[1]{>{\raggedright\arraybackslash}p{#1}}
\newcolumntype{C}[1]{>{\centering\arraybackslash}p{#1}}
\newcolumntype{R}[1]{>{\raggedleft\arraybackslash}p{#1}}

\begin{document}
\preprint{IP/BBSR/2018-18  \newline
 \hspace*{11.9cm} IPPP/18/14}

\title{Probing the Type-II Seesaw Mechanism through the Production of Higgs Bosons at a Lepton Collider}
\author[a,b]{Pankaj Agrawal,}
\author[a,b]{Manimala Mitra,}
\author[c]{Saurabh Niyogi,}
\author[a,b]{Sujay Shil,}
\author[d]{ and Michael Spannowsky}

\affiliation[a]{Institute of Physics, Sachivalaya Marg, Bhubaneswar, Odisha 751005, India}
\affiliation[b]{Homi Bhabha National Institute, Training School Complex, Anushakti Nagar, Mumbai 400085, India}
\affiliation[c]{Gokhale Memorial Girls' College, Kolkata}
\affiliation[d]{Institute for Particle Physics Phenomenology, Durham University, Durham DH1 3LE, United Kingdom}

\emailAdd{agrawal@iopb.res.in}
\emailAdd{manimala@iopb.res.in}
\emailAdd{saurabhphys@gmail.com}
\emailAdd{sujay@iopb.res.in}
\emailAdd{michael.spannowsky@durham.ac.uk}

\abstract{We investigate the production and decays of doubly-charged Higgs bosons for the Type-II seesaw mechanism at an $e^{+} e^{-}$ collider with two center of mass energies, $\sqrt{s}=380$ GeV and 3 TeV, and analyze the fully hadronic final states in detail. Lower mass ranges can be probed during the 380 GeV run of the collider, while high mass ranges, which are beyond the 13 TeV Large Hadron Collider discovery reach, can be probed with $\sqrt{s}=3$ TeV. For such a heavy Higgs boson, the final decay products are collimated, resulting in fat-jets. We perform a substructure analysis to reduce the background and find that a doubly-charged Higgs boson in the mass range 800-1120 GeV can be discovered during the 3 TeV run, with integrated luminosity $\mathcal{L} \sim 95\,   \rm{fb}^{-1}$ of data. For 380 GeV center of mass energy, we find that for the doubly-charged Higgs boson in the range 160-172 GeV, a $5\sigma$ significance can be achieved with only integrated luminosity $\mathcal{L} \sim 24 \, \rm{fb}^{-1}$.  Therefore, a light  Higgs boson can be discovered immediately during the run of a future $e^{+} e^{-}$ collider.}

\maketitle
\section{Introduction \label{intro}}

With the discovery of the Higgs boson at the Large Hadron Collider (LHC), we start to develop an understanding of how the Standard Model (SM) fermion and gauge boson masses are generated in terms of the Brout-Englert-Higgs (BEH) mechanism. However, one of the main puzzles that still remains unclear is the origin of light neutrino masses and mixings. The same BEH mechanism can, in principle be employed to generate Dirac mass of SM neutrinos by extending the SM to include right-handed neutrinos.  However, the required large hierarchy of the Yukawa couplings raises uncomfortable questions. A completely different ansatz is that neutrinos are their own anti-particles and hence, their masses have a different origin than the other SM fermions. A tiny eV Majorana neutrino mass can be generated by the seesaw mechanism,  
where light neutrinos acquire their masses from a lepton number violating (LNV) $d=5$ operator $LLHH/\Lambda$~\cite{Weinberg:1979sa,Wilczek:1979hc}. Such operator is not forbidden as the lepton number is only a classical symmetry of the SM, violated by quantum effects. 

There are three proposed categories, commonly known as, Type-I, -II, and -III seesaw mechanisms in which the SM is extended by a $SU(2)_L$
singlet fermion~\cite{Minkowski:1977sc,Mohapatra:1979ia,Yanagida:1979as,GellMann:1980vs,Schechter:1980gr,Babu:1993qv,Antusch:2001vn}, $SU(2)_L$ triplet scalar boson~\cite{Magg:1980ut,Cheng:1980qt,Lazarides:1980nt,Mohapatra:1980yp}, and $SU(2)_L$ triplet fermion~\cite{Foot:1988aq}, respectively.
In particular, the second possibility, i.e., where a triplet scalar field with the hypercharge $Y = +2$ is added to the SM, is the simplest model with an extended Higgs sector. The neutral component of the triplet acquires a vacuum expectation value (vev) $v_{\Delta}$, 
and generates neutrino masses through the Yukawa interactions. Perhaps, the most appealing feature of this model is its minimality. The same Yukawa interaction between the lepton doublet and the triplet scalar field generates Majorana masses for the neutrinos, and also dictates the phenomenology of the charged Higgs bosons.
 
A number of detailed studies have already been performed at the LHC \cite{Perez:2008ha,Melfo:2011nx,delAguila:2008cj,Chakrabarti:1998qy,Aoki:2011pz,Chun:2013vma, Banerjee:2013hxa} to search for the triplet Higgs. One attractive feature of this model is the presence of the doubly-charged Higgs boson, and its distinguishing decay modes. Depending on the triplet vev, the doubly-charged Higgs boson can decay into same-sign dilepton,  same-sign gauge bosons, or even via a cascade decay \cite{Perez:2008ha,Melfo:2011nx,delAguila:2008cj}. The details of the Higgs spectrum have been discussed in \cite{Arhrib:2011uy,Dev:2013ff}. For the branching ratios and collider signatures, see  \cite{Perez:2008ha,Melfo:2011nx,delAguila:2008cj, Chakrabarti:1998qy,Aoki:2011pz}. The
CMS and ATLAS collaborations have searched for the same-sign dilepton final states for all flavors, and constrained the mass of the doubly-charged Higgs as $M_{H^{\pm \pm}} > 820,  870\, \rm{GeV} $  at  95$\%$ C.L. \cite{Aaboud:2017qph, CMS-PAS-HIG-16-036}. However, this is only relevant for a very tiny vev $v_{\Delta} < 10^{-4}$ GeV, where the doubly-charged Higgs boson decays into the same-sign dilepton with 100$\%$ branching ratio. For larger triplet vev such as $v_{\Delta} \gsim  0.01$ GeV, this branching ratio is negligibly small. Therefore, a direct bound on the mass of the  $H^{\pm \pm}$ from the same-sign dileptonic final state cannot be obtained. An alternative search where the $H^{\pm \pm}$ is produced in association with two jets (vector boson fusion channel)  gives relaxed constraints \cite{Khachatryan:2014sta, Sirunyan:2017ret}. For $v_{\Delta} \ge 10^{-4}$ GeV, the doubly-charged Higgs boson predominantly decays into same-sign diboson. The collider signatures and the discovery prospect of this scenario have been discussed in \cite{Kanemura:2013vxa,Kanemura:2014goa,Kanemura:2014ipa}, and \cite{Mitra:2016wpr, Ghosh:2017pxl} (see \cite{Englert:2016ktc} for the discussion on the composite Higgs model and \cite{Hays:2017ekz} for discussion on flavor violating $\tau$ decays).
Previous searches for $H^{\pm \pm}$ in the pair-production channel and their subsequent decays into same-sign leptons at LEP-II has put a constraint $M_{H^{\pm \pm}} > 97.3 $ GeV at $95 \%$ C.L. \cite{Abdallah:2002qj}.

While a number of searches at the LHC are ongoing to experimentally verify the presence of the doubly-charged Higgs boson, in this work we perform a detailed collider analysis to explore the discovery prospects at a future lepton collider.  For a large mass of the doubly-charged Higgs boson, the pair-production cross-section at the LHC becomes small. Furthermore, the presence of numerous backgrounds weakens its discovery prospects. Therefore, a  lepton collider with a much cleaner environment will be more suitable to search the high mass regime of the doubly-charged Higgs boson. In addition, we also exhaust the low mass regime, yet unconstrained by the LHC, and by LEP-II measurements.  
 
 We consider the pair-production of the doubly-charged Higgs boson at a lepton collider and its subsequent decays into same-sign gauge bosons. We focus on the hadronic decays of the produced gauge bosons and analyze the multi-jet final states in detail. As a prototype example, we consider the future $e^{+}e^{-}$ collider Compact Linear Collider (CLIC) \cite{Battaglia:2004mw,Linssen:2012hp, Abramowicz:2013tzc,AlipourTehrani:2254048}, that will operate with the center of mass energies $\sqrt{s}=380$ GeV, 1.4 TeV  and 3 TeV. We first analyze the discovery reach of the doubly-charged Higgs boson at 380 GeV center of mass energy. Subsequently, we carry out a detailed simulation for the very heavy doubly-charged Higgs boson with a mass around and beyond one TeV. For such a heavy Higgs, its final decay products are collimated,  leading to fat-jets. We perform a jet-substructure analysis and tag the gauge bosons.  We find that a heavy Higgs boson with a mass up to $1120$ GeV can be most optimally discovered with 5$\sigma$ significance at the 3 TeV run of CLIC with 95\,$\rm{fb}^{-1}$ of data. For lower masses, the range $160-172$ GeV can be covered with only  
 $\mathcal{L} \sim 24 \, \rm{fb}^{-1}$ of luminosity.  {For the earlier discussions on Higgs triplet model at a linear collider, see \cite{Shen:2015bna,Blunier:2016peh,Cao:2016hvg,Guo:2016hjt}. For the other SM and BSM searches at CLIC and other linear colliders, see \cite{Contino:2013gna,Heinemeyer:2015qbu,Thamm:2015zwa, Craig:2014una,Durieux:2017rsg, Ellis:2017kfi,Godbole:2011hw,Biswal:2005fh,Abramowicz:2013tzc,Dannheim:2012rn,Ananthanarayan:2014eea, Thomson:2015jda,  Milutinovic-Dumbelovic:2015fba,Wang:2017urv,Abramowicz:2016zbo,Banerjee:2016foh}} for Higgs physics and 
 effective field theory, \cite{Heinemeyer:2016wey,Ari:2013wda,Li:2013zwo, Llamas-Bugarin:2017upv,Chiang:2015rva} for different BSM scenarios, and \cite{Banerjee:2015gca, Antusch:2016ejd,Ahriche:2014xra,Antusch:2015gjw,Antusch:2016vyf,Biswal:2017nfl,Barry:2012ga}
for seesaw and radiative neutrino mass model searches.  For the discussion on probing dark-sector at  $e^{+}e^{-}$ collider,  see \cite{Chacko:2013lna, Andersen:2013rda}. 

Our paper is organized as follows: we briefly review the basics of the Type-II seesaw model in Sec.~\ref{model}. In Sec.~\ref{expcons}, we discuss existing experimental constraints. In the subsequent subsections, Sec.~\ref{modsiglow} and Sec.~\ref{modsighigh}, we analyze in detail the production cross-sections and the discovery potential of the multi-jet final states at the $e^{+}e^{-}$ collider. Finally, we present our conclusions in Sec.~\ref{conclu}.
\section{Model Description {\label{model}}}
In addition to the SM Higgs field $\Phi$, the Type-II seesaw model \cite{Magg:1980ut,Cheng:1980qt,Lazarides:1980nt,Mohapatra:1980yp} contains an additional  $SU(2)_L$ triplet Higgs field
\begin{eqnarray}
 \Delta=\begin{pmatrix} \frac{\Delta^+}{\sqrt{2}} & \Delta^{++} \\ \Delta^0 & -\frac{\Delta^+}{\sqrt{2}}
\end{pmatrix}  ~~~ \sim (1,3,2).
\end{eqnarray}
We denote the neutral components of the SM doublet and triplet Higgs fields as   $\Phi^0=\frac{1}{\sqrt{2}}(\phi^0+i\chi^0)$ and $\Delta^0=\frac{1}{\sqrt{2}}(\delta^0+i\eta^0)$, respectively. The real {scalars} $\phi^0$ and $\delta^0$ acquire vevs denoted as $v_{\Phi}$ and $v_{\Delta}$ with  $v^2=v^2_{\Phi}+v^2_{\Delta}=(246 \, \, \rm{GeV})^2$. The light neutrino mass is proportional to the triplet vev $v_{\Delta}$. The new scalar field $\Delta$, being a triplet under $SU(2)$, interacts with the SM gauge bosons. The relevant kinetic term has the form
\begin{eqnarray}
\mathcal{L}_{\rm{kin}}( \Delta)&=&\rm{Tr}[(D_\mu \Delta)^\dagger (D^\mu \Delta)], 
\label{kinetic}
\end{eqnarray}
with the covariant derivative $D_\mu \Delta=\partial_\mu \Delta+i\frac{g}{2}[\tau^aW_\mu^a,\Delta]+ig'B_\mu\Delta$.
The Yukawa interactions of $\Delta$ with the lepton fields are 
\begin{eqnarray}
\mathcal{L}_Y(\Phi, \Delta)&=& Y_{\Delta}\overline{L_L^{c}}i\tau_2\Delta L_L+\rm{h.c.}.~~~~ 
\label{yukawa}
\end{eqnarray}
In the above,  $Y_{\Delta}$ is a $3\times 3$ matrix and $c$ denotes charge conjugation. The triplet field $\Delta$ 
carries lepton number +2 and hence the Yukawa term conserves lepton number. The scalar potential of the Higgs fields $\Phi$ and $\Delta$ is  
\begin{eqnarray}
V(\Phi,\Delta)&=&m_\Phi^2\Phi^\dagger\Phi+\tilde{M}^2_{\Delta}\rm{Tr}(\Delta^\dagger\Delta)+\left(\mu \Phi^Ti\tau_2\Delta^\dagger \Phi+\rm{h.c.}\right)+\frac{\lambda}{4}(\Phi^\dagger\Phi)^2 \nonumber\\
&+&\lambda_1(\Phi^\dagger\Phi)\rm{Tr}(\Delta^\dagger\Delta)+\lambda_2\left[\rm{Tr}(\Delta^\dagger\Delta)\right]^2 +\lambda_3\rm{Tr}[(\Delta^\dagger\Delta)^2]
+\lambda_4\Phi^\dagger\Delta\Delta^\dagger\Phi,~~~~
\label{eqn:scalpt}
\end{eqnarray}
where $m_{\Phi}$ and $\tilde{M}_{\Delta}$ are real parameters with mass dimension {1}, $\mu$ is the lepton-number violating
parameter with positive mass dimension and $\lambda$, $\lambda_{1-4}$ are dimensionless quartic Higgs couplings.

There are seven physical Higgs states in mass basis, that arise after diagonalization of the scalar mass matrix written in the gauge basis. They are: the charged Higgs bosons $H^{\pm \pm}$, $H^{\pm}$, the neutral Higgs bosons $h^0, H^0 $ and $A^0$. The two charged scalar fields
$\Phi^{\pm}$ of $\Phi$ and $\Delta^{\pm}$ of $\Delta$  mix to give singly-charged states $H^{\pm}$  and the charged Goldstone $\chi^{\pm}$ bosons. Similarly, the mixing between the two CP-odd fields ($\chi^{0}$ and $\eta^{0}$)
gives rise to  $A^0$, and the neutral Goldstone boson $\rho^0$. Finally, we obtain the SM Higgs boson ($h$) and a heavy Higgs boson ($H$) via the mixing of the two neutral CP-even states $\Phi^{0}$ and $\delta^{0}$.

The physical masses of {the} doubly and singly charged Higgs bosons $H^{\pm \pm}$ and $H^{\pm}$  can be written as
\begin{eqnarray}
m_{H^{++}}^2=M_\Delta^2-v_\Delta^2\lambda_3-\frac{\lambda_4}{2}v_\Phi^2,\label{eq:mhpp}~~~
m_{H^+}^2= \left(M_\Delta^2-\frac{\lambda_4}{4}v_\Phi^2\right)\left(1+\frac{2v_\Delta^2}{v_\Phi^2}\right).\label{eq:mhp}
\end{eqnarray}
The CP-even and CP-odd neutral Higgs bosons $h$, and  $H$  have the physical masses
\begin{eqnarray}
m_h^2=\mathcal{T}_{11}^2\cos^2\alpha+\mathcal{T}_{22}^2\sin^2\alpha-\mathcal{T}_{12}^2\sin2\alpha, \label{mh}~
m_H^2=\mathcal{T}_{11}^2\sin^2\alpha+\mathcal{T}_{22}^2\cos^2\alpha+\mathcal{T}_{12}^2\sin2\alpha.\label{mH}
\end{eqnarray}
In the above $\mathcal{T}_{11}$,  $\mathcal{T}_{22}$ and $\mathcal{T}_{12}$ have the following expressions: 
 \begin{eqnarray}
\mathcal{T}_{11}^2=\frac{v_\Phi^2\lambda}{2},~~
\mathcal{T}_{22}^2=M_\Delta^2+2v_\Delta^2(\lambda_2+\lambda_3), ~~
\mathcal{T}_{12}^2=-\frac{2v_\Delta}{v_\Phi}M_\Delta^2+v_\Phi v_\Delta(\lambda_1+\lambda_4).
\end{eqnarray}
The CP-odd Higgs field $A^0$ has the mass term
\begin{eqnarray}
m_A^2 &= &M_\Delta^2\left(1+\frac{4v_\Delta^2}{v_\Phi^2}\right) \label{mA},\quad \mathrm{with}~~ M^2_{\Delta}=\frac{v^2_{\Phi} \mu}{\sqrt{2} v_{\Delta}}.
\end{eqnarray}

{The  difference between $H^{\pm \pm}$ and $H^{\pm}$ masses  is dictated by the coupling $\lambda_4$ of the scalar potential.  For a positive $\lambda_4$, the ${H^{\pm \pm}}$ is lighter than ${H^{\pm }}
$.} The mass difference $\Delta M^2$ is 
\begin{equation}
\Delta M^2=M^2_{H^{\pm}}-M^2_{H^{\pm \pm}} \sim \frac{\lambda_4}{2} v^2_{\Phi}+\mathcal{O}(v^2_{\Delta}).
\label{diffchdmass}
\end{equation}
Throughout our analysis, we consider the mass hierarchy $M_{H^{\pm \pm}} < M_{H^{\pm}}$.  

Due to the non-trivial representations of $\Delta$, the Higgs triplet has interactions with a number of SM fermions and gauge bosons. This opens up a number of possible decay modes that can be explored at the LHC, and at future linear colliders. In the next section, we summarize the different direct experimental constraints on the doubly-charged Higgs boson mass and triplet vev.

\begin{figure}[t]
\begin{center}
\includegraphics[scale=0.5]{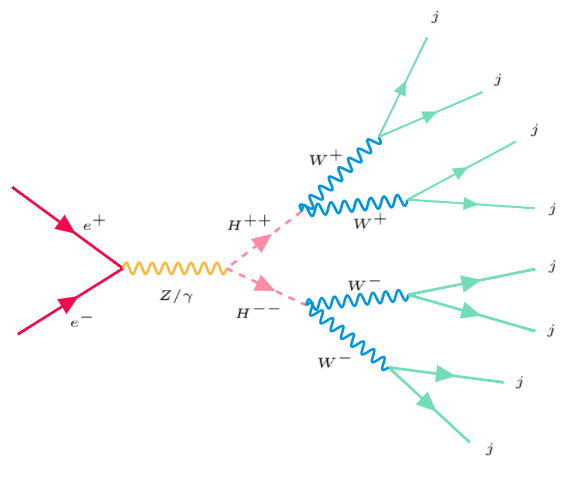}
\caption{The Feynman diagram for $H^{++} H^{--}$ pair-production and its subsequent decays into gauge bosons.} 
\label{f:feyndiag380}
\end{center}
\end{figure}

\section{Decay Modes and Experimental Constraints \label{expcons}}

The most characteristic feature of the Type II seesaw model is the presence of the doubly-charged Higgs boson $H^{\pm \pm }$, that can decay into the leptonic or bosonic states and gives unique signatures at high energy colliders. 
The different decay modes and the branching ratios of the $H^{\pm \pm }$ depend on the triplet vev $v_{\Delta}$.  For smaller triplet vev, the $H^{\pm \pm}$ predominantly decays into the same-sign leptonic states $H^{\pm \pm } \to l^{\pm} l^{\pm}$, whereas for  larger $v_{\Delta}$, the gauge boson mode $H^{\pm \pm} \to W^{\pm} W^{\pm}$ becomes dominant \cite{Perez:2008ha, Melfo:2011nx}.  The relevant decay widths are calculated to

\begin{equation}
\Gamma (H^{\pm \pm} \to l^{\pm}_i l^{\pm} _j)=\Gamma_{l_i l _j}=\frac{M_H^{\pm \pm} } {(1+\delta_{ij}) 8 \pi}   \left |\frac{M_{\nu_{ij}}}{v_{\Delta}} \right |^2, \, \, M_{\nu}=Y_{\Delta} v_{\Delta},
\end{equation}

\begin{equation}
\Gamma (H^{\pm \pm} \to W^{\pm} W^{\pm})=\Gamma_{W^{\pm }W^{\pm }}=\frac{g^2 v^2_{\Delta}}{8 \pi M_{H^{\pm \pm}}} \sqrt{1- \frac{4}{r^2_W}} \left[ \left (2+(r_W/2-1)^2 \right ) \right ].
\end{equation}
In the above $M_{\nu}$ denotes the neutrino mass matrix, $i,j$ are the generation indices,  $\Gamma_{l_i l_j}$ and $\Gamma_{W^{\pm} W^{\pm}}$ are the partial decay widths for the $H^{\pm \pm} \to l^{\pm}_i l^{\pm}_j$, and $H^{\pm \pm} \to W^{\pm} W^{\pm}$  channels, respectively. The parameter  $r_W$ denotes  the ratio of $H^{\pm \pm}$ and the $W$ gauge boson masses, $r_W=\frac{M_{H^{\pm \pm}}}{M_W}$.  The branching fraction of the leptonic and bosonic mode becomes equal around the triplet vev $v_{\Delta } \sim 
10^{-4}$ GeV \cite{Perez:2008ha, Melfo:2011nx}.

A number of searches have been proposed at the LHC to discover   $H^{\pm \pm}$ using multilepton signatures. The  searched modes in \cite{ Perez:2008ha, Melfo:2011nx, delAguila:2008cj, Mitra:2016wpr} are pair and associated production with the $H^{\pm\pm}$ decaying into leptonic or gauge boson states.  Below we discuss the existing constraints on  $H^{\pm \pm}$ from LEP and LHC searches.

\begin{itemize}

\item

{{ Constraint from LEP-II}}: The search for doubly-charged Higgs boson $H^{\pm \pm}$ decaying into charged leptons have been performed at LEP-II. This constrains the mass parameter $M_{H^{\pm \pm}} > 97.3$ GeV  \cite{Abdallah:2002qj} at 95$\%$ C.L.

\item

{Constraints from pair and associated production}:  Stringent constraint on the $M_{H^{\pm \pm}}$ by analyzing $H^{\pm \pm } \to l^{\pm } l^{\pm}$ have been placed at the 13 TeV LHC. The CMS collaboration looked for different leptonic flavors including $e e, e \mu, e \tau, \mu \mu, \mu \tau $ 
and $\tau \tau$. In addition, the CMS searches also include the associated production $p p  \to H^{\pm \pm} H^{\mp}$ and the subsequent decays, $H^{\pm} \to l^{\pm} \nu$. This combined channel of pair-production and associated production gives the stringent constraint $M_{H^{\pm \pm}} > 820$ GeV \cite{CMS-PAS-HIG-16-036} 
at $95 \%$ C.L for $e, \mu$ flavor.  The constraint from ATLAS searches comes from pair-production. The bound is $M_{H^{\pm \pm}} > 870 $ GeV at $95 \%$ C.L \cite{Aaboud:2017qph}. 
Note that these limits are valid only for a small triplet vev $v_{\Delta} < 10^{-4}$ GeV. 

\item

Constraint from VBF: For larger values of the triplet vev {$v_{\Delta} \ge 10^{-4}$ GeV}, the leptonic branching ratio {becomes smaller}. Instead the {decay} mode  $H^{\pm \pm} \to W^{\pm } W^{\pm}$ is dominant. Therefore the searches in vector boson fusion (VBF) become more important. A search for $p p  \to j j H^{\pm \pm} \to j j  W^{\pm } W^{\pm}$ at the 8 TeV LHC in the VBF channel sets a constraint on the triplet vev $v_{\Delta} \sim 25$ GeV for $M_{H^{\pm \pm} } \sim 300$ GeV \cite{Khachatryan:2014sta}. This constraint has been updated  \cite{Sirunyan:2017ret} using 13 TeV data at the LHC.  

\end{itemize}

\begin{figure}[t]
\begin{center}
\includegraphics[scale=0.85]{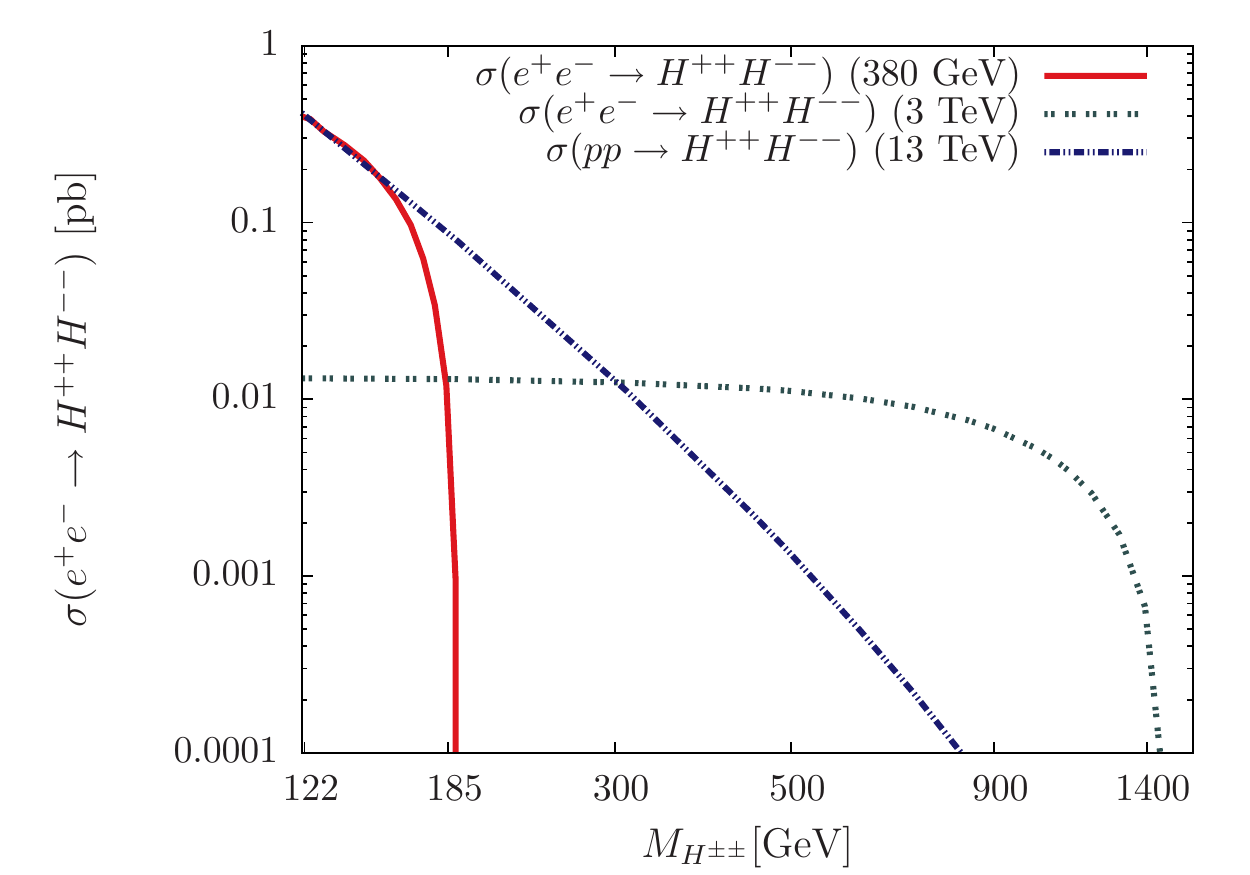}
\caption{The production cross-section at $e^{+} e^{-}$ collider.  The center of mass energies are $\sqrt{s}=380$ GeV and $3$ TeV.  For comparison, we also show the cross-section at 13 TeV LHC.  The pair-production cross-section 
increases by a factor of two, if CLIC uses 80$\%$, and 30$\%$ beam polarization for electron and positron beam.  }
\label{f:feyndiag2}
\end{center}
\end{figure}

Note that, for extremely small $v_{\Delta}$, the mass of the doubly-charged Higgs  boson is very tightly constrained from pair-production searches. For a larger triplet vev, this constraint significantly relaxes. The VBF cross-section scales quadratically with the triplet vev and hence, increases for a very large vev. However, the range of $v_{\Delta} \sim 10^{-4}-10^{-1}$ GeV cannot be probed at the 13 TeV LHC in {VBF channel}, as the   cross-section becomes extremely small in this range. Recently, in \cite{Ghosh:2017pxl}, the authors have looked for pair-production of $H^{\pm \pm}$ in  large $v_{\Delta}$ region and analyzed the signature where the final state contains  di-lepton,    multi-jet, and missing energy. The lighter mass $M_{H^{\pm \pm} } \lsim 190$ GeV can be probed at the 14 TeV LHC with 3000 $\rm{fb}^{-1}$ of data. {In \cite{Kanemura:2014ipa}, the authors have used LHC 8 TeV run-I result of same-sign di-lepton to derive a bound $M_{H^{\pm \pm}} \ge 84 $ GeV, relevant for large $v_{\Delta}$. } {For large  mass of the doubly-charged Higgs, the LHC cross-section however becomes significantly smaller, as shown in Fig.~\ref{f:feyndiag2}. On the other hand, the fall in the cross-section at a $e^{+}e^{-}$ collider is relatively  smaller.} This motivates us to explore the signatures of {doubly-charged Higgs} at a  {lepton} collider, where the cross-section still remains larger for heavy charged Higgs masses. In the following sections, we explore the scope of a future linear collider to probe large $v_{\Delta}$ region with a) a very low mass range of $H^{\pm \pm }$, that is still experimentally allowed, and b) a very heavy highly boosted $H^{\pm \pm}$.

\section{Large triplet vev and collider signatures}

In this section, we analyze the collider signatures of a doubly-charged Higgs boson at an $e^{+}e^{-}$ collider and explore the sensitivity reach to probe low and high mass regimes. Throughout our analysis, we  consider a large triplet vev $v_{\Delta} \ge 10^{-4}$ GeV, where the present experimental constraints are weak. As the prototype example, we consider CLIC \cite{Battaglia:2004mw,Linssen:2012hp, Abramowicz:2013tzc,AlipourTehrani:2254048} that will operate with three different center of mass energies $\sqrt{s}=380, 1400$ GeV and $3$ TeV. We present our simulation for $380$ GeV and $3 $ TeV center of mass energies.  The doubly-charged Higgs boson, $H^{\pm \pm}$, can be produced at  $e^{+}e^{-}$ collider via photon and $Z$-boson mediated diagrams, as shown in Fig.~\ref{f:feyndiag380}. We show in Fig.~\ref{f:feyndiag2} the respective production cross-sections. As both of the diagrams are s-channel processes, the cross-section reduces with increasing center of mass energy. For a relatively small center of mass energy $\sqrt{s}=380$ GeV, the maximum 
cross-section reaches up to $\sigma \sim 506$ fb for $M_{H^{\pm \pm}}=102$ GeV. A rapid decline in the cross-section occurs near $M_{H^{\pm \pm}} \sim 190$ GeV,  close to  kinematical threshold.  
For the choice of large $v_{\Delta}$, the produced particles  $H^{\pm \pm}$ will decay into $W^{\pm }W^{\pm}$ gauge bosons with almost 100$\%$ branching ratio.  In the following, we will first discuss the low-mass regime, that can be probed in the $\sqrt{s}=380$ GeV run. Following that we discuss the high-mass regime, that can be explored at 3 TeV center of mass energy and gives rise to specific signatures of boosted Higgs boson. In both cases we focus on multi-jet final states.

\subsection{Low mass $H^{\pm \pm}$ at $\sqrt{s}={380}$ GeV  \label{modsiglow}}

We consider the pair-production of  $H^{\pm \pm}$, and its subsequent decay into $W^{\pm}W^{\pm}$ at $\sqrt{s}=380$  GeV.  The produced $W^{\pm}$ decay dominantly into hadronic final states. Thus, to retain as much signal rate as possible, we focus on fully hadronic channel. Therefore, our model signature comprises of multi-jet events. In the subsequent analysis, we demand a high jet multiplicity, i.e., the number of jets  $N_{\rm{jet}} \ge 7$. For the signal, the production processes are

\begin{itemize}

\item
 $e^{+}  e^{-} \to H^{\pm \pm} H^{\mp \mp} \to 4W  \ge 7j$ for $M_{H^{\pm \pm} } \gsim \ 2 M_W$
 \item
 $e^{+}  e^{-} \to H^{\pm \pm} H^{\mp \mp} \to W^{\pm} j j W^{\mp} j j  \ge 7j$ for $M_{H^{\pm \pm} } < \ 2 M_W$

\end{itemize}

\begin{table}[!ht]
\centering
\begin{tabular}{||c|c|c|c||}
\hline 
\multicolumn{4}{|c||}{ $e^+ e^-  \to H^{++} H^{--}  \to N_j \ge 7j $}\\ \hline \hline
Mass (GeV)  & $ \sigma_p$ (fb)  & $\sigma_d( N_j \ge 7 j)$  (fb)  &  $\sigma_d( N_j \ge 7 j+b\,  \rm{veto})$  (fb) \\
\hline
121   &  0.80 &  0.30 & 0.20  \\
137 & 2.08 &  0.94 & 0.66   \\
159 & 5.45&  2.58 & 1.82 \\
172 & 5.04 &  2.48 & 1.74\\
184 & 1.11 &  0.53 & 0.38 \\
\hline \hline
\multicolumn{4}{|c||}{Backgrounds} \\ \hline \hline
Processes  & $ \sigma_p$ (fb) $\times 10^{-2}$  & $\sigma_d( N_j \ge 7 j)$ (fb)$\times 10^{-2}$  & $\sigma_d( N_j \ge 7 j+b\,  \rm{veto})$  (fb) $\times 10^{-2}$\\
\hline
$e^{+} e^{-} \to t \bar{t} \rightarrow 6j $ & 10341.0 & 338.0  &  36.0\\
$W^{+}W^{-}3j, W^{\pm} \rightarrow 2j$ & 8.89 &  1.18 & 0.88 \\
$ZZ+3j, Z \rightarrow 2j$ & 0.98 &    0.13 & 0.10 \\
$ 7j $ & 30.32 &  1.13 & 0.88  \\
$W^{\pm}+5j, W^{\pm} \rightarrow j j$ & 30.18 &  4.64 & 3.54\\
$Z+5j, Z \rightarrow j j$ & 18.32 &  2.15 & 1.61 \\
\hline 
\end{tabular} 
\caption{The cross-sections  for the signal and background for the fully hadronic final states,  arising from $e^{+} e^{-} \to H^{\pm \pm} H^{\mp \mp}$.  $\sigma_p$ refers to the partonic cross-section. $\sigma_d$ is the cross-section after taking into account  detector effects. The last column represents the cross-section with  $b$-veto. The center of mass energy is  $\sqrt{s}=380$ GeV and kinematic cuts are specified in the text.
}
\label{tab:7jlowmass}
\end{table}

In the former scenario the $H^{\pm \pm}$ decays predominantly into on-shell $W^{\pm} W^{\pm}$, while in the latter case $H^{\pm \pm}$ decays into one on-shell and one off-shell gauge bosons with subsequent decays into jets.

To simulate the events, we use  first  {\tt FeynRules} \cite{Alloul:2013bka} and  generate the  model file via Universal Feynrules Output (UFO) \cite{Degrande:2011ua, deAquino:2011ub}. We  compute the hard processes with the package {\tt MadGraph5\_aMC@NLO}  \cite{Alwall:2014hca},  and pass the output (in LHE format) through {\tt Pythia 6} \cite{Sjostrand:2001yu}  for showering and hadronization.  The detector simulation has been taken into account by {\tt Delphes-3.3.0} \cite{deFavereau:2013fsa}, where we use the ILD card. Here we use anti-$k_{t}$ jet clustering algorithm \cite{Cacciari:2008gp} to form jets. Similar final states will be generated from a number of SM processes.  
We consider the following sets of backgrounds  and perform a detailed simulation:

\begin{itemize}
\item
$e^{+} e^{-} \to t \bar{t} \to 6 j $

\item
$e^{+} e^{-} \to W^{+} W^{-} +3 j , W^{\pm} \to  2 j$, and $e^{+} e^{-} \to ZZ+3 j, Z  \to 2 j$

\item
$e^{+} e^{-} \to 7 j $  

\item
$e^{+} e^{-} \to W^{\pm}+5 j, W^{\pm} \to  j j$, and $e^{+} e^{-} \to Z+5 j, Z \to  j j$

\end{itemize}

Among the backgrounds,  $e^{+} e^{-} \to 7j$ includes diagrams of coupling order $\alpha_{EW}^{2}\alpha_{S}^{5}$ with quarks and gluons as intermediate particles. As listed above, we treat the $t \bar{t}$ and gauge boson mediated backgrounds separately. 
For the partonic event generation, we implement the following sets of cuts at MadGraph level both for the signal and backgrounds: the transverse momentum of  light jets $p_T(j_i) > 20$ GeV for all the final state partons, the pseudo-rapidity $|\eta| < 5.0$, and the separation between the light jets $\Delta R(j_i, j_j) > 0.4$. 

We consider few illustrative mass points between $M_{H^{\pm \pm}} \sim 121$ GeV and the kinematic threshold $M_{H^{\pm \pm}} \sim 184$ GeV, and display the signal cross-sections in Table.~\ref{tab:7jlowmass}. The cross-sections $\sigma_p$ refers to the partonic cross-section, while $\sigma_d$ is after taking into account reconstruction and detector effects.  In addition to the cuts at the partonic level, we further implement few more selection cuts: the transverse momentum of jets $p_T(j_i) > $ 20 GeV for all the jets, pseudo-rapidity $|\eta| < 4.5$ for jets,  and the number of jets $N_j \ge 7j$. The largest background arises from $t \bar{t} \to 6j$, where the cross-section is about $103$ fb at the partonic level. {This is much larger   than the largest signal cross-section} {$5.45$} fb, corresponding to $M_{H^{\pm \pm}}=159$ GeV. For other mass points, the ratio is even bigger.   However, demanding high jet multiplicity $N_{j}\ge 7j$ reduces this background to $\sigma_d \sim {3}$ fb. For the masses of the doubly-charged Higgs boson $M_{H^{\pm \pm}}=159$ and $ 172$ GeV, the signal and background cross-sections become almost equal after demanding higher jet multiplicity.  A few comments are in order:

\begin{table}[!ht]
\centering
\begin{tabular}{||c|c|c||}
\hline 
\multicolumn{3}{|c||}{ $e^+ e^-  \to  H^{++} H^{--}  \to N_j \ge 7j  $}\\ \hline \hline
Mass (GeV)  & $n_s $ & $\mathcal{L} \, (\rm{fb}^{-1})$  \\
\hline
121   & 1.54 & 1054.14  \\
137 & 4.48  & 124.56 \\
159 &  10.48  & 22.76  \\
172 &  10.15 & 24.26 \\
184 &  2.65  & 355.99  \\
\hline \hline
\end{tabular} 
\caption{ The statistical significance $n_s$ for $\mathcal{L}=100 $ $\rm{fb}^{-1}$. The third column displays the required luminosity to achieve 5$\sigma$ significance. The center of mass energy is $\sqrt{s}=380$ GeV.}
\label{tab:380gevnsig}
\end{table}

\begin{table}[!b]
\centering
\begin{tabular}{||c|c|c||}
\hline 
\multicolumn{3}{|c||}{ $e^+ e^-  \to  H^{++} H^{--}  \to N_j \ge 7j+ b\,  \rm{veto}  $}\\ \hline \hline
Mass (GeV)  & $n_s(b) $ & $\mathcal{L} \, (\rm{fb}^{-1})$  \\
\hline
121   & 2.52  & 393.67  \\
137 & 6.33  & 62.39\\
159 & 12.14   &   16.96\\
172 & 11.84  & 17.83\\
184 & 4.23  &  139.72\\
\hline \hline
\end{tabular} 
\caption{ The statistical significance $n_s(b)$ for $\mathcal{L}=100 $ $\rm{fb}^{-1}$ and the required luminosity to achieve 5$\sigma$ significance, after implementing the $b$-veto.  The  center of mass energy is  $\sqrt{s}=380$ GeV. }
\label{tab:380gevnsigbtag}
\end{table}

\begin{itemize}

\item
Between the higher and lower mass ranges, i.e., $M_{H^{\pm \pm}} > 2 M_W$ and $M_{H^{\pm \pm}} < 2 M_W$, the former scenario corresponds to larger {pair-production} cross-sections. 
The fall in cross-section in the higher mass range occurs when $M_{H^{\pm \pm}} \sim {184} $ GeV, where it approaches the kinematic threshold.  For lower mass ranges, $M_{H^{\pm \pm}}  \sim 121$ GeV, the reduction of cross-section {after the detector effect} occurs due to stronger kinematic cuts. 
The produced jets from a $H^{\pm \pm}$ with mass $M_{H^{\pm \pm}}  \sim $ 121 GeV are often quite soft. With the constraint on jet transverse momentum  $p_T > 20 $ GeV the reconstruction efficiency becomes smaller. 

\item

The signal comprises of hadronic final states with higher jet multiplicity. For the signal,  $H^{\pm \pm}$ decays into two $W^{\pm}$ with subsequent decay into quarks, resulting in a final state with $N_j=8$. At a $e^{+} e^{-}$ collider, there are only a few SM processes that can generate a similar final state. A full reconstruction of the signal results in a fairly low reconstruction efficiency. Thus, we allow for one jet to be too soft or out of the kinematic cuts range.

\end{itemize}

In Table.~\ref{tab:380gevnsig}, we derive the statistical significance $n_s=\sigma_d(S) \sqrt{\mathcal{L}}/\sqrt{\sigma_d(S)+\sigma_d(B)}$ for our benchmark points corresponding to Table.~\ref{tab:7jlowmass}. Here $\sigma_d(S)$ and $\sigma_d(B)$ represent the final cross-sections for the signal and background after all the selection cuts. Additionally, we also show the required luminosity to achieve a 5$\sigma$ significance. {Other than the extreme low and high mass ranges $M_{H^{\pm \pm}}=121$ and $ 184 $ GeV, all other mass points have a  large discovery prospect with 124 $\rm{fb}^{-1}$ of data}. In particular,  {we show that  the doubly-charged Higgs boson with intermediate masses of 159 GeV (172 GeV) can be discovered with 5$\sigma$ significance with only $\mathcal{L} \sim 22~(24) \, \rm{fb}^{-1}$, respectively. This further improves to $\mathcal{L} \sim 16 ~(17) \, \rm{fb}^{-1}$ after applying a $b$-veto ($50-60$\% efficiency and $1$\% miss-tag efficiency), that helps in reducing the dominant 
top-quark pair background.


\subsection{Boosted Heavy  $H^{\pm \pm}$ at $\sqrt{s}={3}$ TeV \label{modsighigh}}

We now consider heavy $H^{\pm \pm}$ with a mass $M_{H^{\pm \pm}} \sim 1$ TeV and its decay into {like-sign} $W^{\pm }W^{\pm}$ gauge bosons. The produced  $W^{\pm}$ decays into hadronic as well as leptonic states. As before, we focus on the purely hadronic final states, which has the largest branching ratio. For such heavy $H^{\pm \pm} $, each of the produced $W^{\pm \pm}$ boson will have large transverse momentum. For a 1.1 TeV $H^{\pm \pm}$, their transverse momentum peaks around $p_T \sim $ 1 TeV, and most of the $W^{\pm }$ are produced in the central region. We show the transverse momentum, and  the pseudo-rapidity distribution of $H^{\pm \pm}$  in Fig.~\ref{f:hpppteta}, for the illustrative benchmark points  $M_{H^{\pm \pm}}=800$ GeV, 1120 GeV and 1.4 TeV.

\begin{figure}[t]
\begin{center}
\includegraphics[scale=0.551]{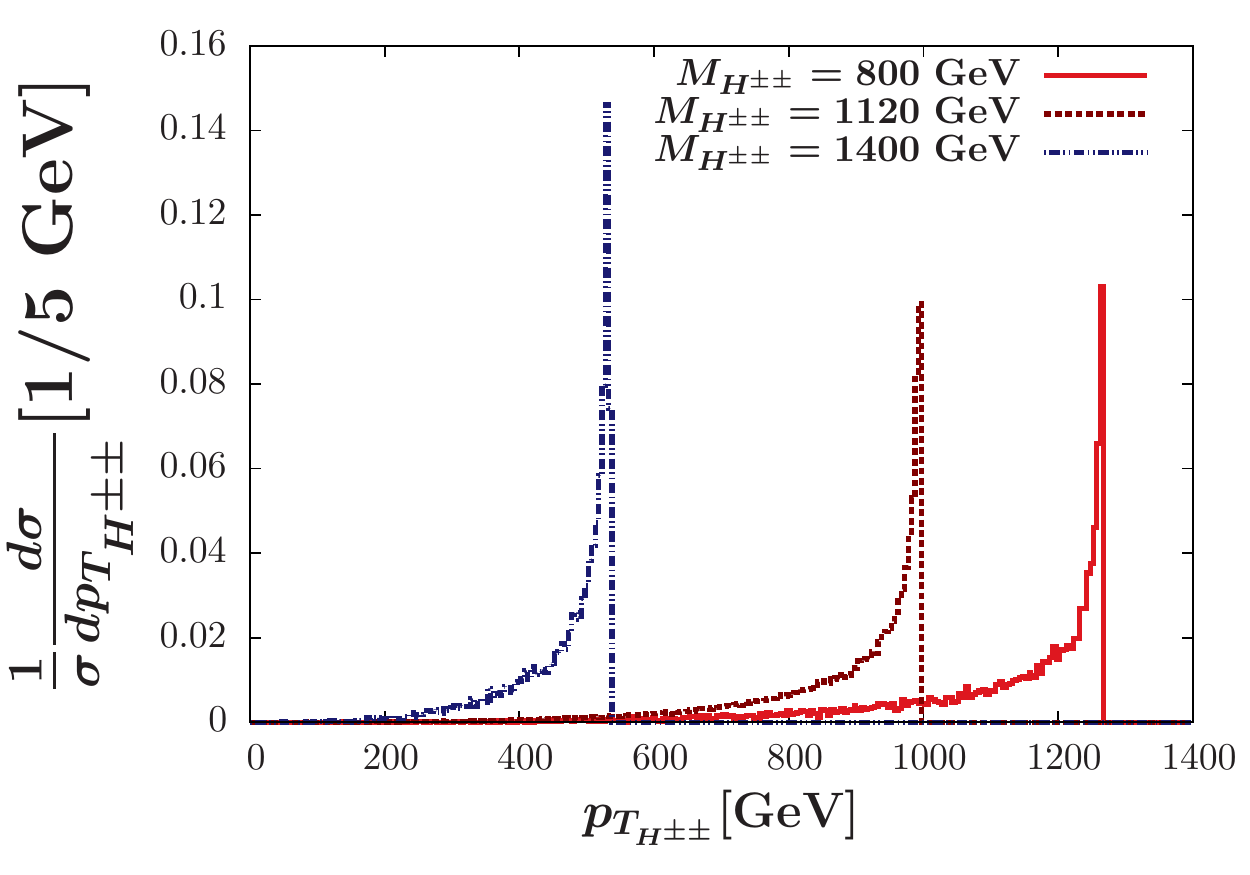}
\includegraphics[scale=0.551]{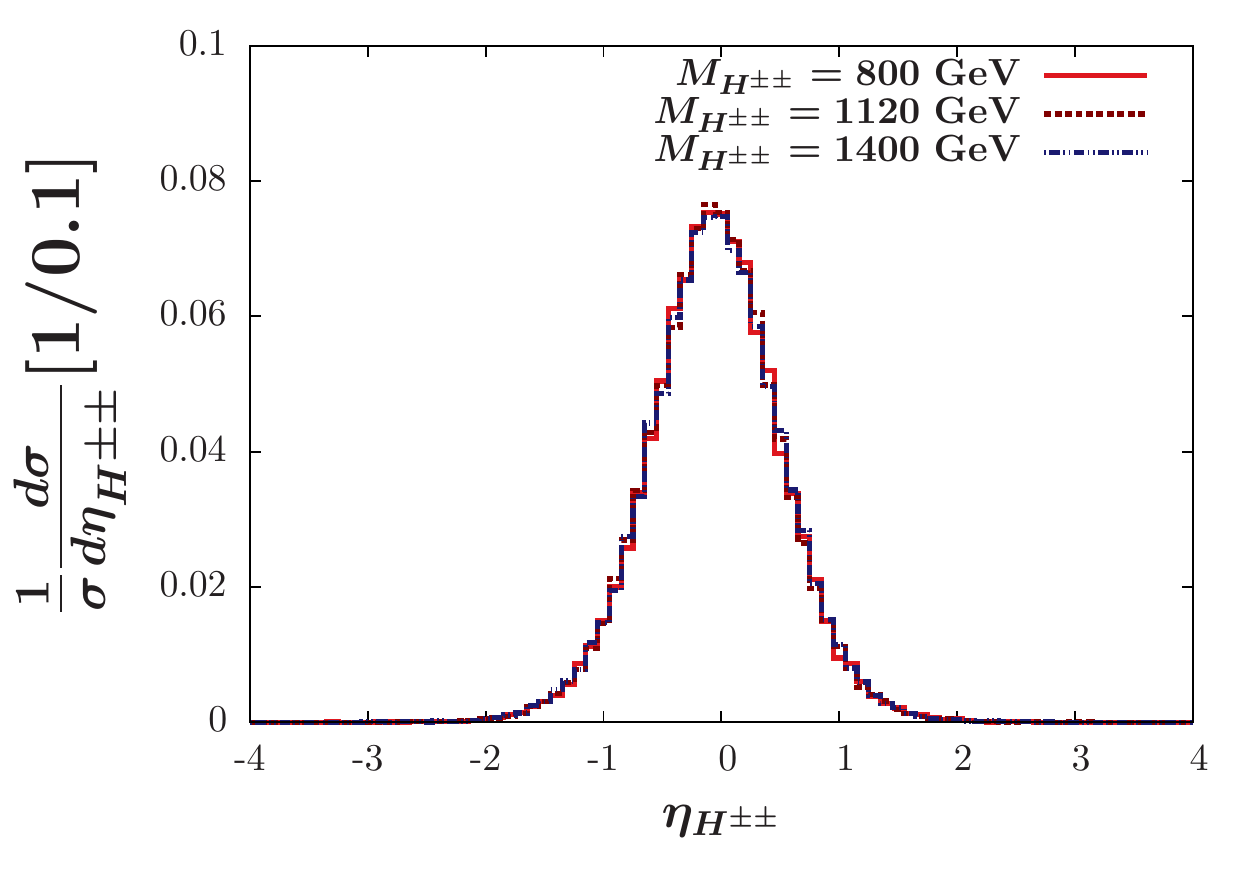}
\caption{The normalized distribution of the transverse momentum and the pseudo-rapidity for the produced $H^{\pm \pm }$. } 
\label{f:hpppteta}
\end{center}
\end{figure}

\begin{figure}[t]
\begin{center}
\includegraphics[scale=0.4]{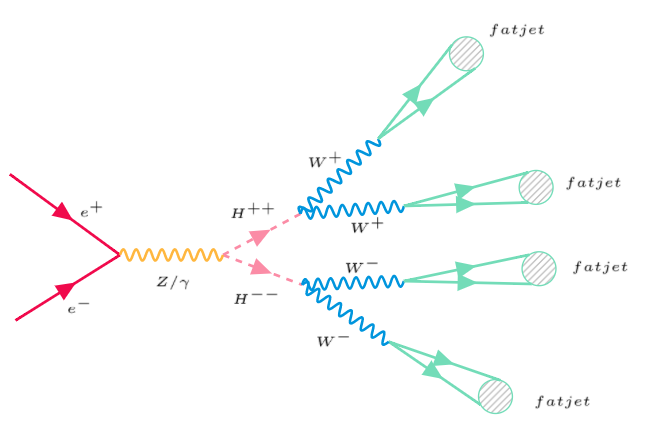}
\caption{The Feynman diagram for $H^{++} H^{--}$ pair-production and its subsequent decays to 4 fat-jet.} 
\label{f:feyndiag3}
\end{center}
\end{figure}

The  final decay products of such heavy Higgs bosons are highly collimated, and can be reconstructed as fat-jets, see Fig.~\ref{f:feyndiag3}.  Therefore, our model signature for such high mass $H^{\pm \pm}$ is

\begin{itemize} 
\item
$e^{+} e^{-} \to H^{\pm \pm} H^{\mp \mp} \to W^{\pm} W^{\mp}W^{\pm} W^{\mp} \to  4\,  \mathrm{fat-jet}$. 
\end{itemize}

\begin{figure}[b]
\begin{center}
  \includegraphics[scale=0.55]{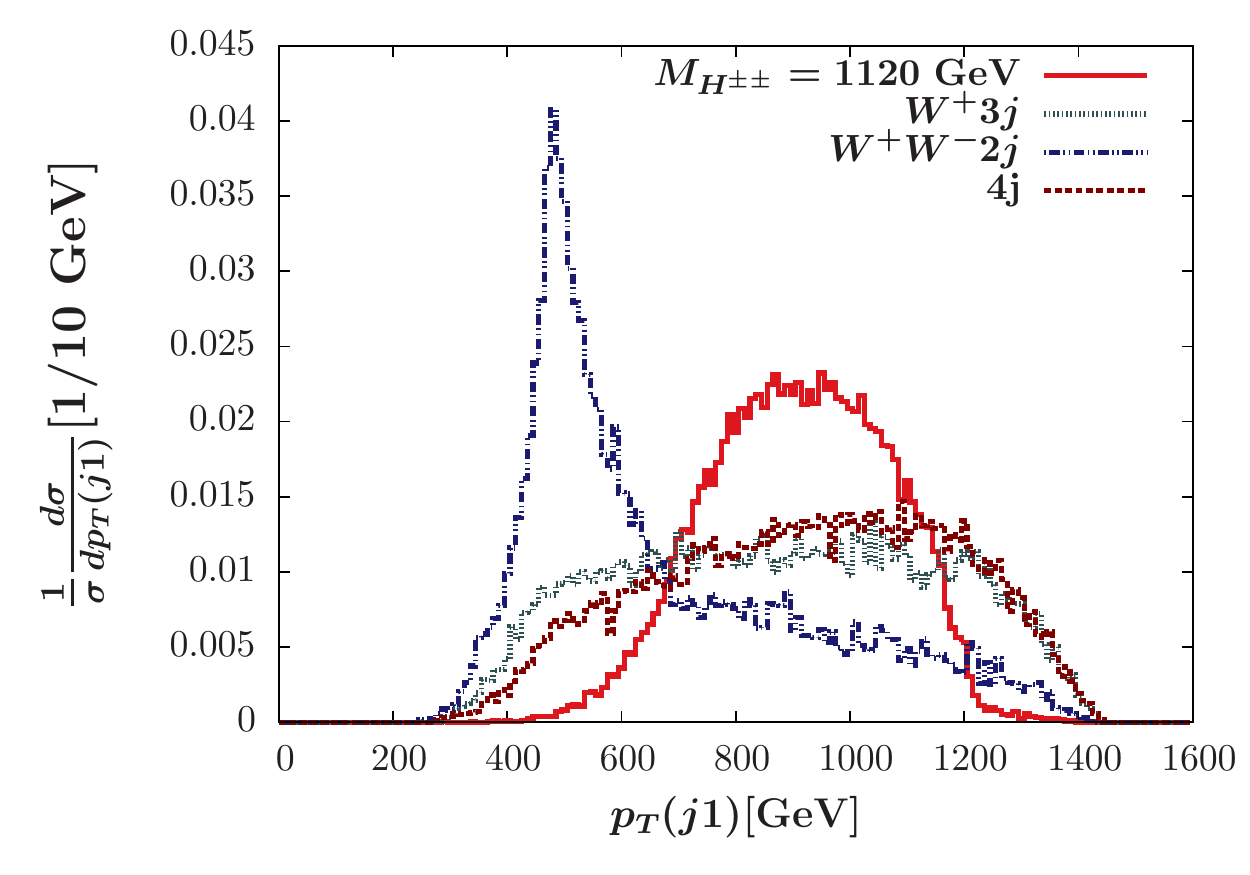}
  \includegraphics[scale=0.55]{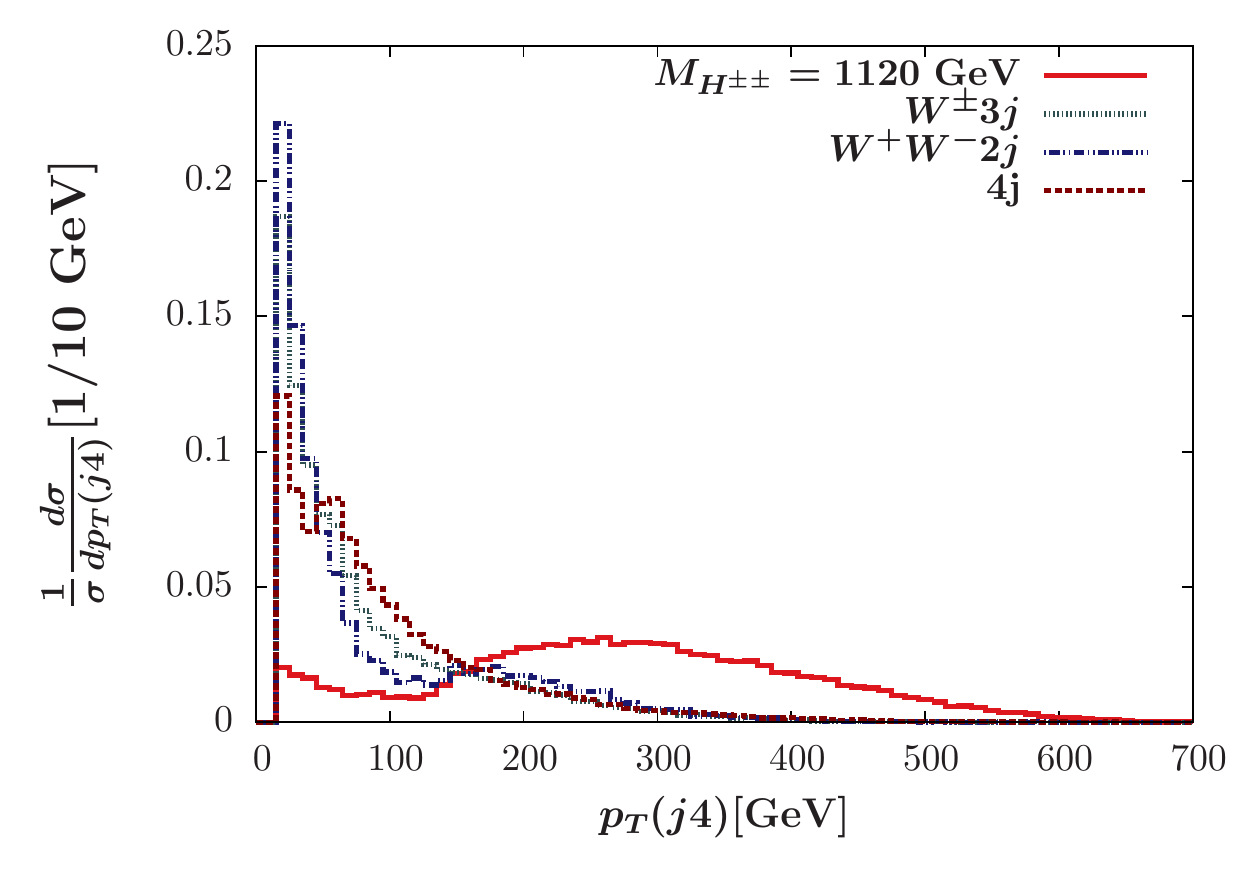}
\caption{The $p_T$ distribution of the leading and 4th leading fat-jets. For signal, we consider $M_{H^{\pm \pm}}=1120$ GeV.}
\label{f:ptfj}
\end{center}
\end{figure}

To generate signal and backgrounds we use the same tool-chain as in Sec.~\ref{modsiglow} except the use of Delphes. Here we analyze the output of {\tt Pythia 8} \cite{Sjostrand:2007gs} (in {\tt HepMC} \cite{Dobbs:2001ck} format)and recluster fat-jets using  Cambridge-Achen algorithm \cite{Dokshitzer:1997in} in {\tt FastJet-3.0.0}\cite{Cacciari:2011ma} with radius parameter $R=1.0$. In Fig.~\ref{f:ptfj}, we show the transverse momentum of the leading fat-jet $j_1$ and the 4th leading fat-jet $j_4$. A number of backgrounds can lead to the final states with multiple fat-jets. These are:  $4j$ (includes both the QED and QCD contributions),  $W^{+}W^{-} 2j$, and  $W^{+}/W^{-}3j$, $W^{+}W^{-} Z j j$ and $t \bar{t}$, with subsequent decays of $W$ boson and the top quark into jets.  The partonic cross-sections of the signal and backgrounds are shown in Table.~\ref{tab:cf4jfat}. The cross-sections for $W^{+}W^{-}Z j j$ and  $t \bar{t}$ are small compared to other backgrounds. Therefore, we do not include these backgrounds in our final analysis.  
Below we discuss in detail the pre-selection and selection cuts for the  signal and backgrounds:

\begin{table}[!ht]
\centering
\small\addtolength{\tabcolsep}{-5pt}
\begin{tabular}{||c|c|c|c|c|c|c|c||}
\hline 
\multicolumn{8}{|c||}{ $e^+ e^-  \to H^{++} H^{--} \to W^{+} W^{+} W^{-} W^{-} \to N j_{\rm{fat}}$} \\ \hline \hline
Masses (GeV)  & $ \sigma_p$ (ab)& $4j_{\rm{fat}}$ ($>120$ GeV)  & 4 MD   &  1 tagged  & 2 tagged & 3-tagged & 4-tagged \\
\hline
800   &  1250 & 812.9 & 758.0 & 757.9 & 748.9 & 671.8 & 389.0 \\
1000 & 850.6 & 527.0 & 492.5 & 492.3 & 486.1 & 436.6 & 258.9 \\
1120 & 670.0 & 380.0 & 358.4 & 358.3 & 354.2 & 321.9 & 193.1 \\
1350 & 167.1 & 80.4 & 75.54 & 75.52 & 74.88 & 68.2 & 42.0 \\
1400 & 94.36 & 45.54 & 42.85 & 42.84 & 42.42 & 38.6 & 24.0 \\
\hline \hline
\multicolumn{8}{|c||}{Backgrounds} \\ \hline \hline
Processes  & $ \sigma_p$ (ab)  & $4j$ ($>120$ GeV)  & 4 MD  & 1 tagged & 2 tagged & 3-tagged & 4-tagged \\
\hline
$4j$ & 6900.0 & 1310.0 & 895.0 & 360.0 & 68.0 & 5.5 & 0.0 \\
$W^{+}3j$ \& $W^{-}3j$ & 1900.0 & 320.0 & 220.0 & 166.0 & 44.0 & 4.8 & $1.52\times 10^{-1}$\\
$W^{+}W^{-}2j$ & 190.0 & 25.6 & 17.7 & 15.6 & 8.3 & 1.23 & $5.7 \times 10^{-2}$\\
$W^{+}W^{-}Z j j$ & 4.23 & -  & -  & - & - & - & - \\
$ t \bar{t} $ & 42  & - & - & - &  - & -  & - \\
\hline 
\end{tabular} 
\caption{The cut-flow for the signal and backgrounds. The cross-sections are in fb. $\sigma_p$ refers to the partonic cross-section. In the backgrounds the decays of the $W^{\pm}$ boson and top quark to jets are included. Here MD refers to Mass-drop. See text for details.
}
\label{tab:cf4jfat}
\end{table}

\begin{figure}[!b]
\begin{center}
  \includegraphics[scale=0.550]{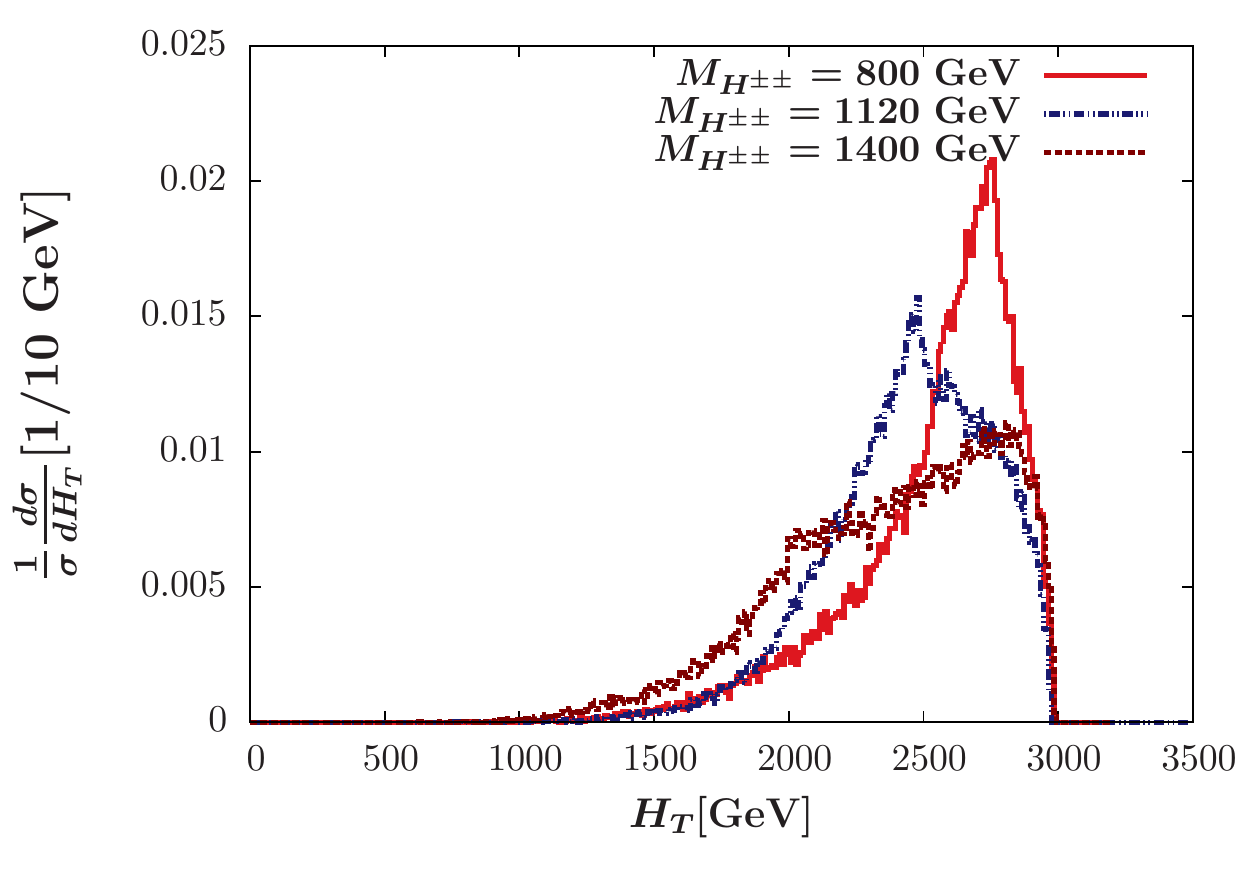}
  \includegraphics[scale=0.55]{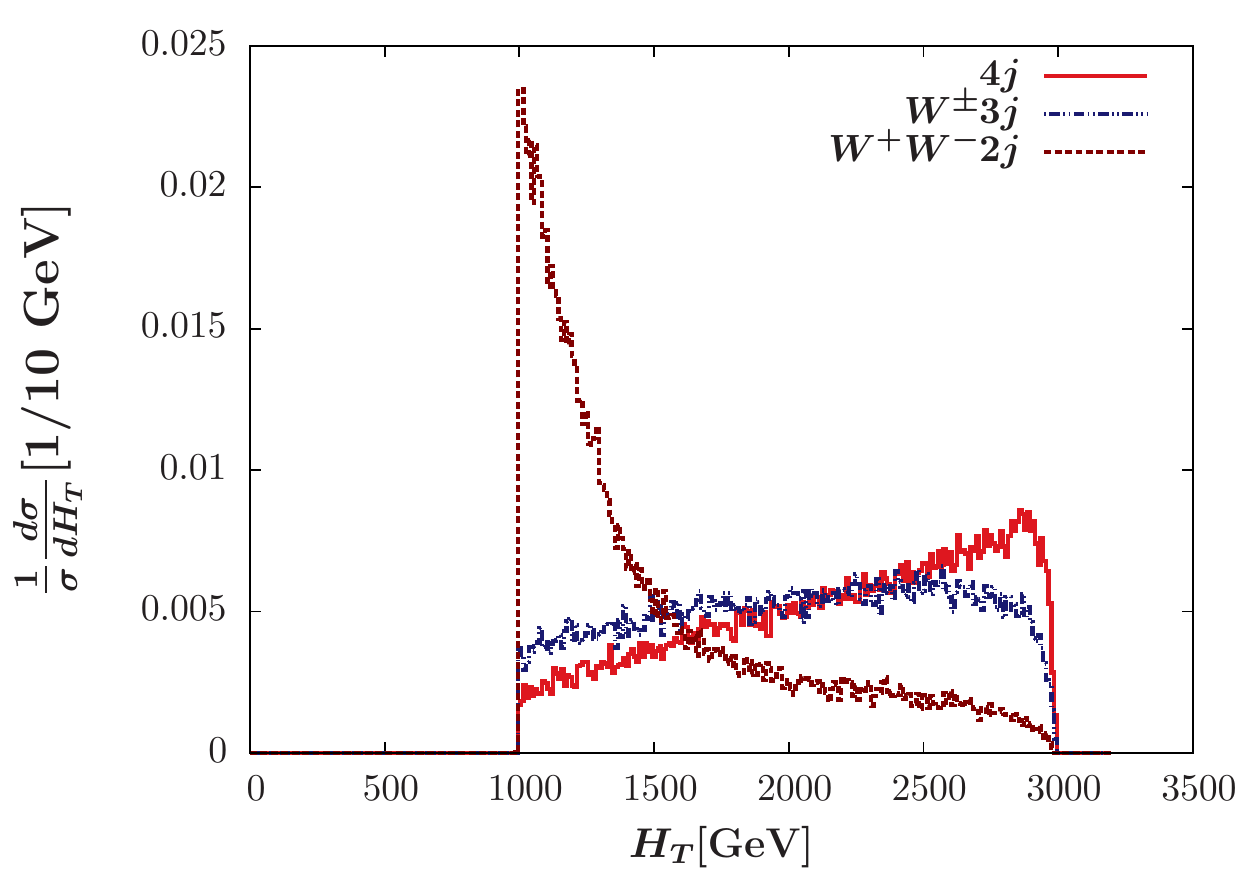}
\caption{The $H_T$ distribution of the jets at the partonic level. We consider three illustrative benchmark points  $M_{H^{\pm \pm }}=$800, 1120, and 1400 GeV. %
}
\label{f:HThpp}
\end{center}
\end{figure}

\begin{itemize}

\item  Most of the signal events are in the central region with pseudo-rapidity distributed around $\eta_{H^{\pm \pm}} \sim 0$, as can be seen in Fig.~\ref{f:hpppteta}.  Additionally, the signal jets have a very high $H_T$(scalar sum of transverse momentum of all final state particles), as shown in Fig.~\ref{f:HThpp}. We consider no cuts on the signal at the parton level.  While generating the backgrounds, we consider the following partonic cuts for $4j$ -  the transverse momentum 
of the jets $p_T > 60$ GeV, and  the jet-jet separation $\Delta R(j,j) > 0.6$; for {\bf{$W^{+}W^{-}2j (W^{\pm}> 2j$)}} and {\bf{$W^{+}3j (W^{\pm}> 2j$)}}- $p_T > 60$ GeV for the leading 4-jets,  {the transverse momentum $p_T > 20$ GeV 
for the remaining  jets,} and  the jet-jet separation $\Delta R(j,j) > 0.4$. The $H_T$ and pseudo-rapidity cut is the same for all the backgrounds, $H_T > 1000$ GeV and $|\eta|< 2.5$. For $t\bar{t}$ samples we have put $\Delta R ({b,j})>0.4$ separation cut and transverse momentum cut on leading two light jet as $p_{T} > 60.0$ GeV. Additionally, we also demand $p_{T}$ of the bottom quarks more than 60 GeV and the  $p_{T}$ of the remaining light quarks more than $20$ GeV.

\item The $\Delta R$ separation of the produced {$W^{+}W^{+}$, $W^{+}W^{-}$} are shown in Fig.~\ref{f:drjjww}. It is evident that for relatively lower masses of $H^{\pm \pm}$, such as 800 GeV, the $W^{+}$ and $ W^{+}$  are closer, as compared to 1400 GeV. This occurs as the  $H^{\pm \pm}$ with 800 GeV mass is more boosted than the higher mass $H^{\pm \pm}$. Hence,  the produced $W^{\pm \pm}$ are more collimated.

\begin{figure}[t]
\begin{center}
  \includegraphics[scale=0.551]{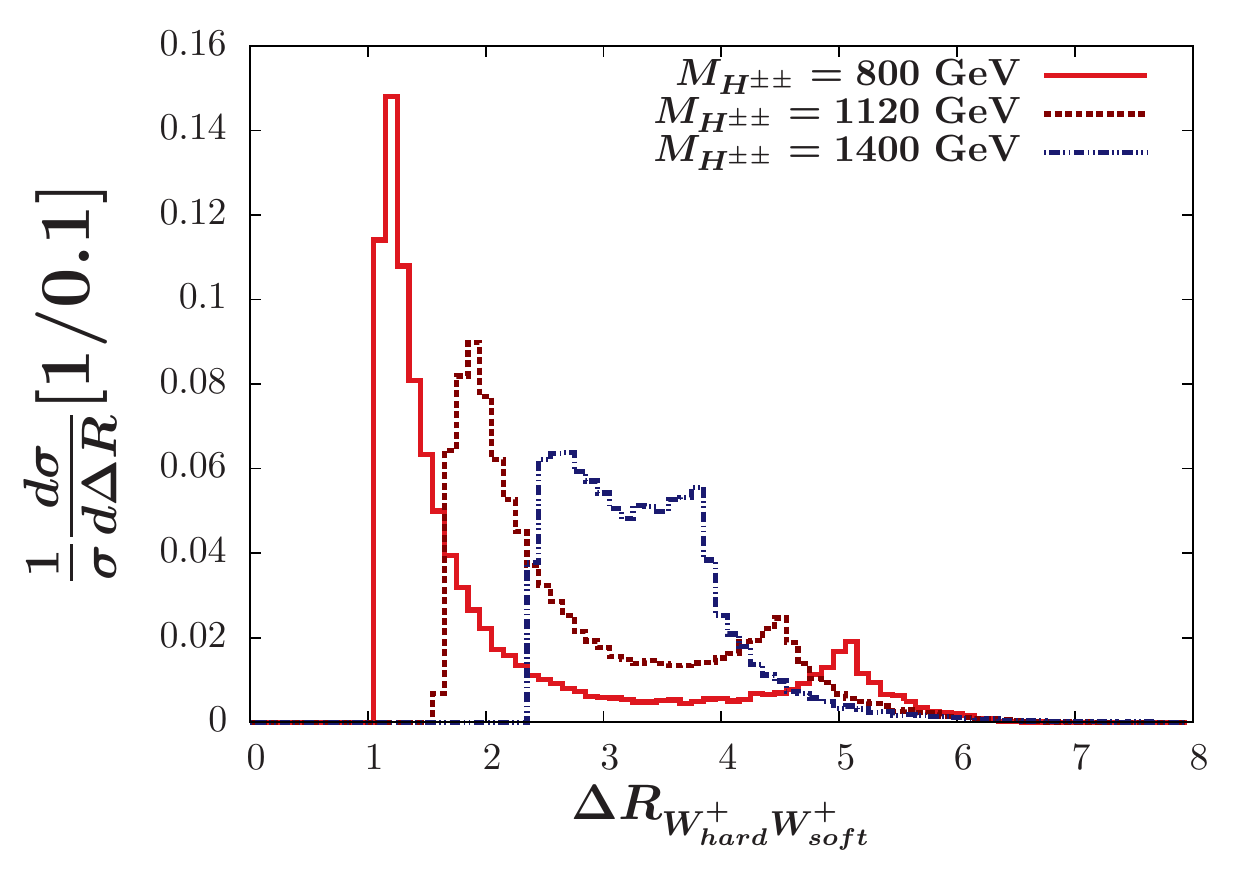}
  \includegraphics[scale=0.551]{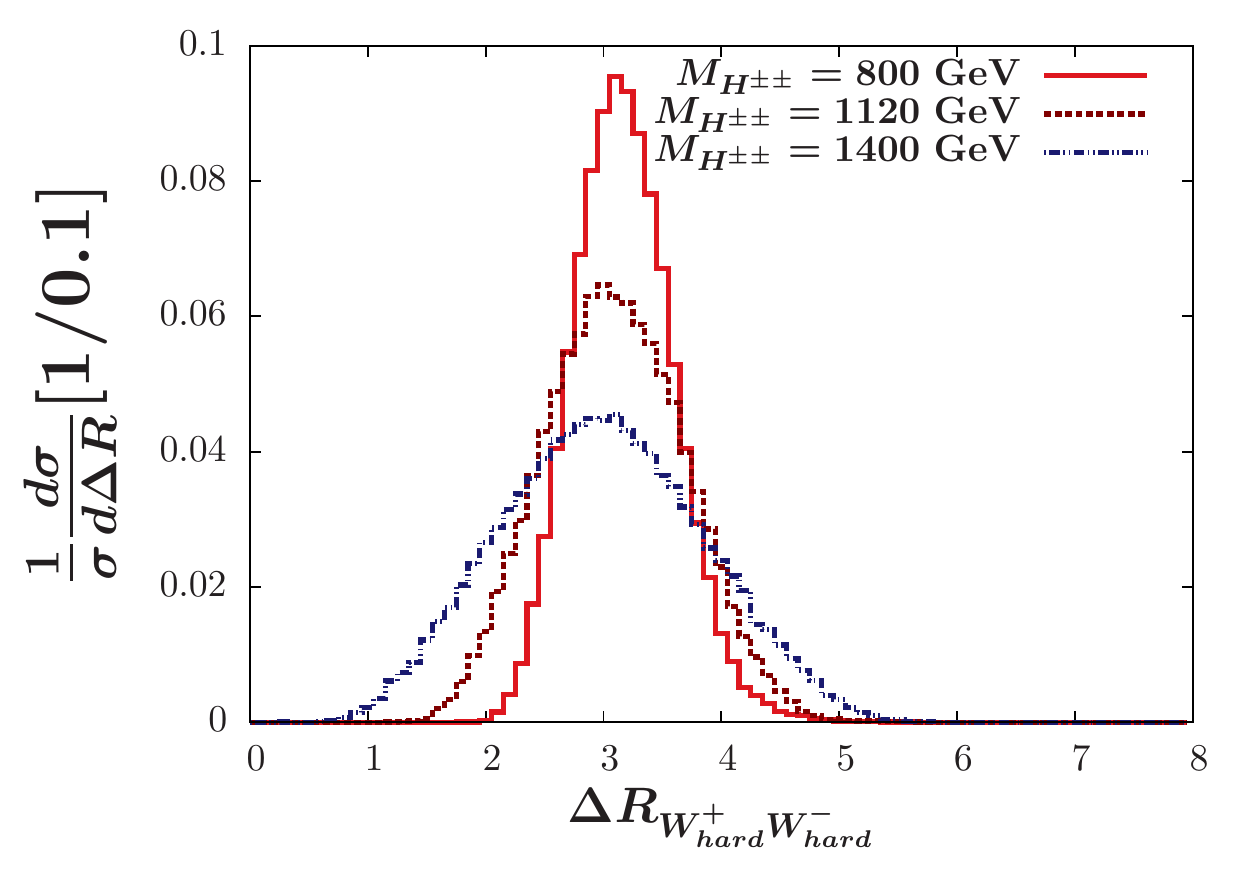}
  \includegraphics[scale=0.551]{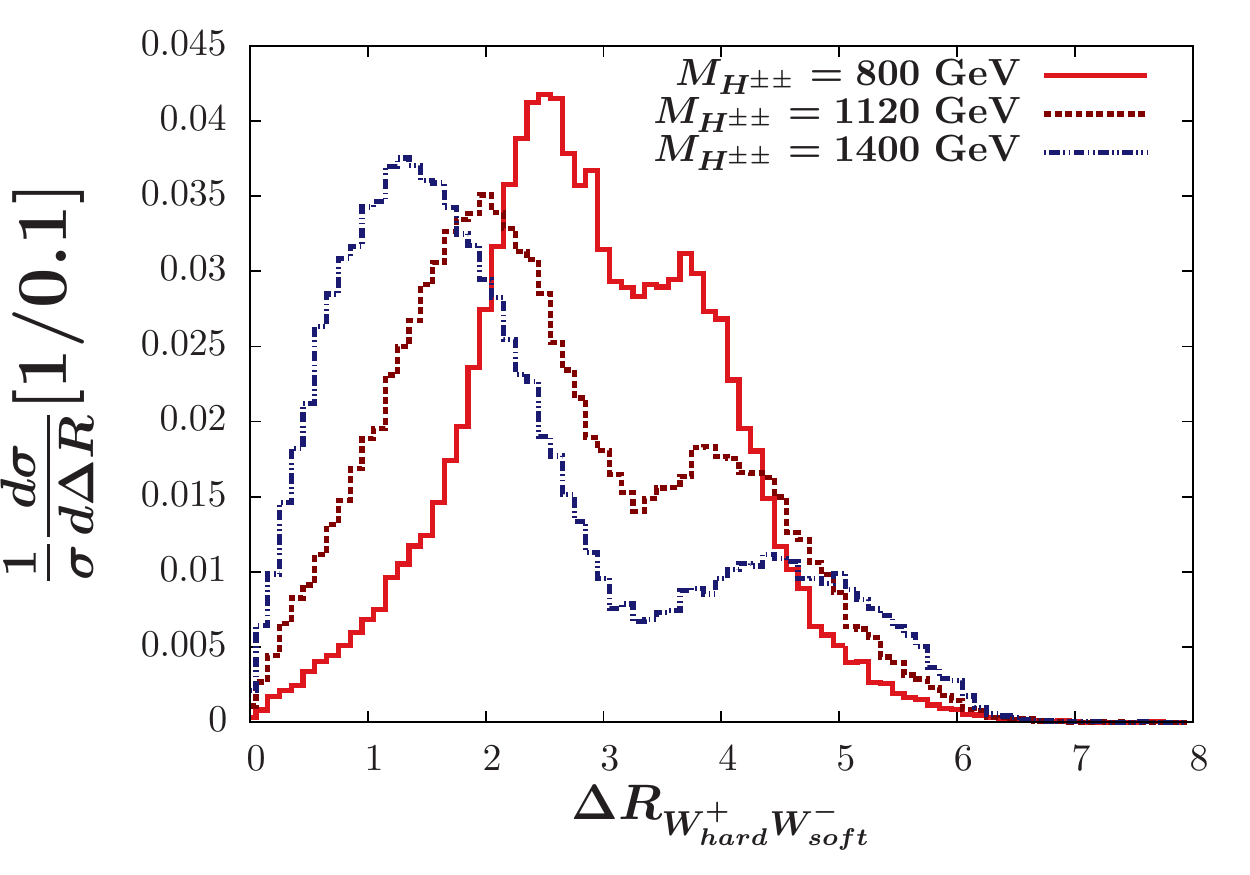}
  \includegraphics[scale=0.551]{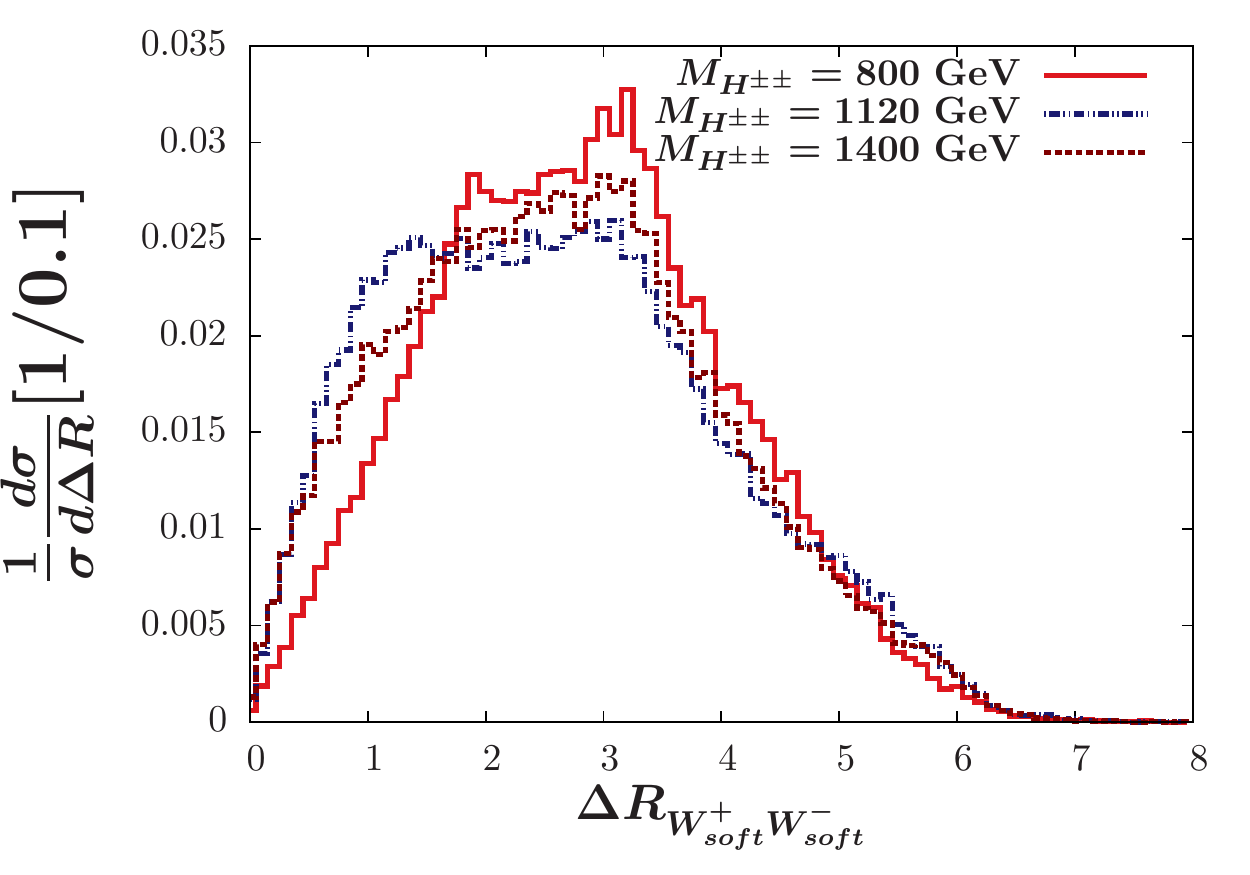}
\caption{The $\Delta {R_{WW}}$ distribution of the different $W^{\pm}$, produced from the doubly charge Higgs $H^{\pm\pm}$. The $W$'s in this figure are $p_{T}$ ordered.}
\label{f:drjjww}
\end{center}
\end{figure}

\item The model signature  contains four fat-jet with high momentum.  We show in Fig.~\ref{f:ptfj}, the transverse momentum of leading and 4th leading fat-jet for $M_{H^{\pm \pm}}=1120$ GeV. Additionally, we also show the distributions of the backgrounds. It is evident that most of the jets have larger transverse momentum for signal, with $p_T \gg 100$ GeV.  Therefore, we design our selection cuts as a) the number of fat-jets $N_{j_{\rm{fat}}}=4$, b) $p_{T_{j_{\rm{fat}}}} > 120$ GeV for all the fat-jets. 

\item We further carry out substructure analysis for the fat-jets. To reconstruct the $W$ bosons we use the mass-drop tagger \cite{Butterworth:2008iy} of which compares the energy-sharings of subjets to indicate if the fat-jet was initiated by a $W$ boson or a parton. For the signal and background, we show the invariant mass of the two sub-jets inside the fat-jet in Fig.~\ref {f:invmfj}. For the signal, the subjets inside a fatjet are generated from the $W$. Therefore, the distribution peaks around the $W$ mass. For the different backgrounds, $4j$ gives flat distribution, while $W^{+}W^{-}2j$ and $W^{+}3j$ shows smaller peak
around $M_W$. As shown in Table.~\ref{tab:cf4jfat}, the backgrounds are significantly reduced after applying the selection cut $|M_{j_1 j_2} - M_W| \le 20$ GeV. 
Here, $M_{j_1 j_2}$ is the invariant mass of the subjets $j_1$ and $j_2$ inside a fat-jet. A detailed cut-flow chart is given in Table.~\ref{tab:cf4jfat}. If at least one fat-jet passes
the invariant mass selection cut, we have 1-tagged event; if at least two fat-jet pass the cut,
we have 2-tagged event and so on.

\end{itemize}

From Table.~\ref{tab:cf4jfat}, the effect of the substructure analysis is clearly evident. The largest background arises from the $e^{+} e^{-} \to 4j$ events. At the partonic level we find a cross-section of $\sigma_p(4j) \sim 6.9\,  \rm{fb} \gg \sigma_p(signal)$. The higher transverse momentum cut on jet $p_T$ reduces the signal nominally, and the background by more than  $\mathcal{O}(5)$ for $4j, WW2j $ and $W+3j$. Demanding that $4$ fat-jets have a non-trivial substructure (referred to as mass-drop MD in  Table.~\ref{tab:cf4jfat}) reduces the background even more. Finally, with the invariant mass cut for the subjets, all backgrounds become almost negligible. For the $H^{\pm \pm}$ masses between 800 GeV to 1.1 TeV one can achieve a $S/B  \sim \mathcal{O}(10)$.  We show the required luminosity to achieve a discovery in Table.~\ref{tab:3tevnsig}. The 800-1120 GeV doubly-charged Higgs boson can be discovered with 39 - 95 $\rm{fb}^{-1}$ of data with at least 2 fat-jet tagged as W-bosons.  However, for higher masses, such as 1.4 TeV a minimum 3 tagged jets will be required.

\begin{figure}[t]
\begin{center}
  \includegraphics[scale=0.55]{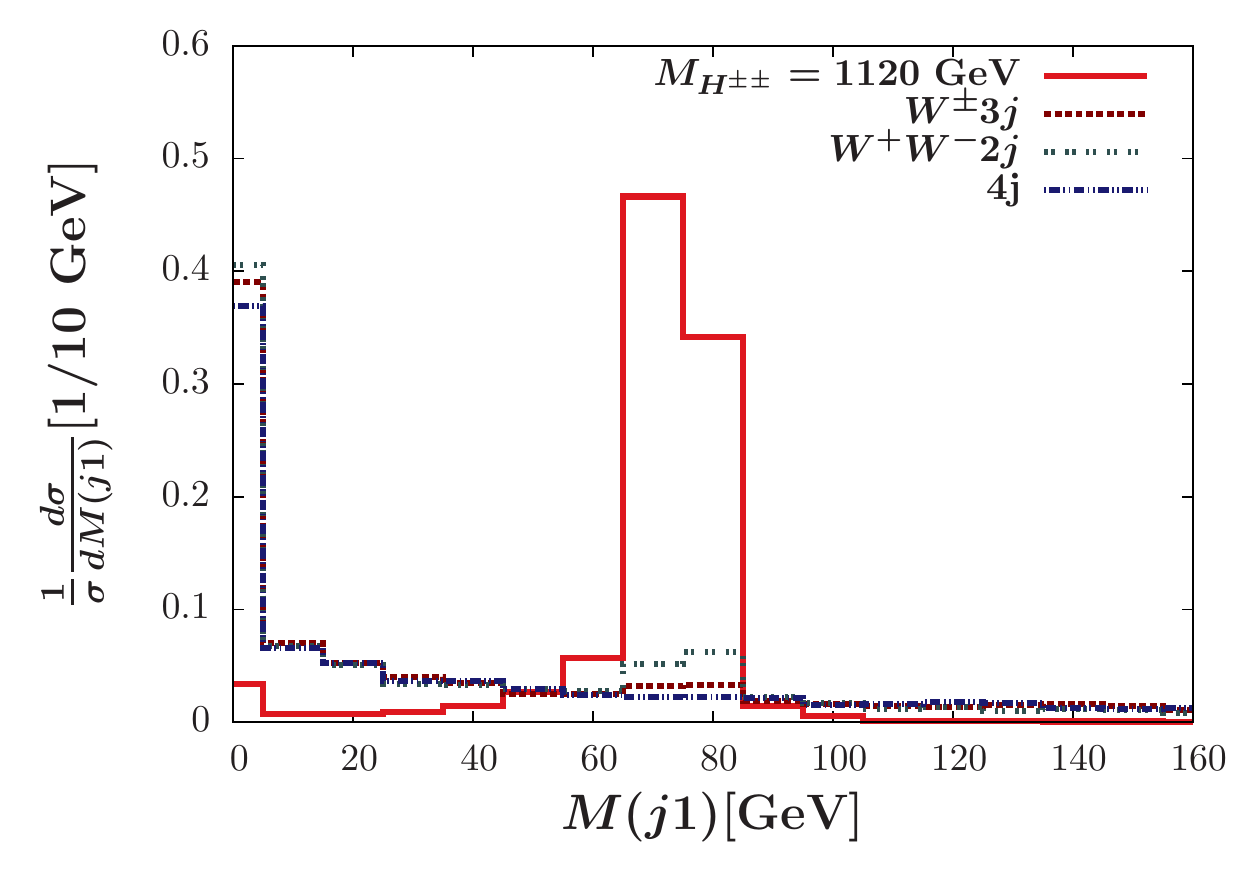}
  \includegraphics[scale=0.55]{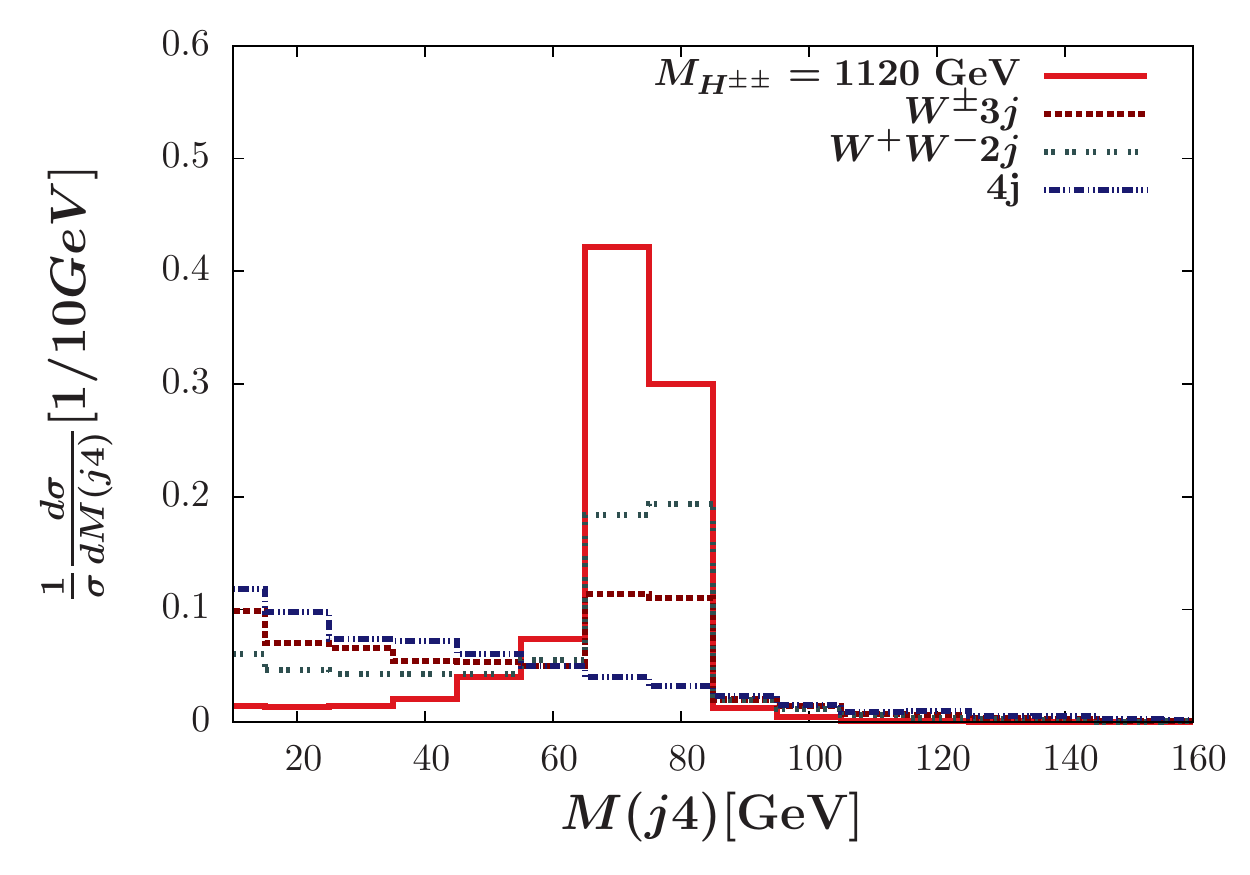}
\caption{{The invariant mass of the fat-jet(leading and 4th leading) constructed using sub-jets four momentum. For signal, we consider $M_{H^{\pm \pm}}=1120$ GeV.}}
\label{f:invmfj}
\end{center}
\end{figure}

\begin{table}[!ht]
\centering
\begin{tabular}{||c|c|c||}
\hline 
\multicolumn{3}{|c||}{ $e^+ e^-  \to H^{++} H^{--}  \to  W^{+} W^{+} W^{-} W^{-} \to N j_{\rm{fat}} $}\\ \hline \hline
Masses (GeV)  & $n_s$ ($2,  3$-tagged $\mathcal{L}=500 \,\rm{fb}^{-1}$ )  &  $\mathcal{L}(\textrm{fb}^{-1})$( with  2,3-tagged)   \\
\hline
800   & 17.96(2-tag)  & 38.75   \\
1000 & 13.95(2-tag)  & 64.23 \\
1120 &  11.49(2-tag) &   94.68\\
1350 & 5.40(3-tag) &  428.66\\
1400 & 3.85(3-tag) &    843.31\\
\hline \hline
\end{tabular} 
\caption{ The statistical significance $n_s$ for $\mathcal{L}=500$ $\rm{fb}^{-1}$ and the required luminosity to achieve 5$\sigma$ significance. The  c.m.energy is  $\sqrt{s}=3$ TeV. In the 2nd column, to derive significance, we consider 2 tagged events for 800-1120 GeV mass range and 3 tagged events for the higher mass range. Here 2-tag implies two or more than two fat-jet masses are {within the window of 60-100 GeV, and the fat-jets are tagged as $W$ jets. Similar criteria applies for 3-tagged jets.}
}
\label{tab:3tevnsig}
\end{table}

\section{Discussion and Conclusions \label{conclu}}

The Type-II seesaw model consists of an extension of the scalar sector by a Higgs triplet field $\Delta$ with hypercharge $Y= +2$.  The neutral component of the triplet acquires a vev and generates the light neutrino mass. 
One of the most attractive features of this model is the presence of the doubly-charged Higgs boson $H^{\pm \pm}$.  Depending on the triplet vev, $H^{\pm \pm}$ can decay into a number of final states,  including same-sign leptons, same sign gauge bosons, and via cascade decay to three body final states. {For the lower triplet vev where $H^{\pm \pm} \to l^{\pm} l^{\pm}$ decays are predominant, the doubly-charged Higgs boson mass  is tightly constrained by LHC pair and {associated} production searches, $M_{H^{\pm \pm}} > 820, 870 $ GeV. However, the higher triplet vev region is poorly constrained by the VBF searches}.   Moreover, the LHC search is limited in the very high mass region $M_{H^{\pm \pm}} \sim 1 $ TeV, where the cross-section is tiny.

In this work, we consider an $e^{+} e^{-}$ collider operating with two center of mass energies $\sqrt{s}=380$ GeV and  3 TeV, and probe the large $v_{\Delta}$ region $v_{\Delta} \ge 10^{-2}$ GeV.  We consider two mass regimes,  a) light $H^{\pm \pm}$ with mass $M_{H^{\pm \pm}}  \lsim 180 $ GeV,  and b) a very heavy $H^{\pm \pm}$ with mass $M_{H^{\pm \pm}} \sim 800-1400 $ GeV.  We consider fully hadronic decays of the produced $W$'s  and perform a detailed analysis for the multi-jet final states.  

For the 380 GeV  center of mass energy, we look into multi-jet final states with $N_{j} \ge 7j$.  We find that a doubly-charged Higgs boson with mass $M_{H^{\pm \pm}} \sim 160-172 $ GeV can be discovered in the immediate run of the $e^{+} e^{-}$ collider, with only  integrated luminosity $\mathcal{L} \sim 24 \, \rm{fb}^{-1}$.  This improves considerably once we apply a $b$-veto, reducing the $t \bar{t}$ background to $\sigma \sim \mathcal{O}(0.1)$ fb. 

The higher mass range $M_{H^{\pm \pm}} \ge $ 1 TeV  can be probed in the $\sqrt{s}=3$ TeV run of the $e^{+}  e^{-}$ collider.  Note that, for such high masses of $H^{\pm \pm}$ the pair-production cross-section at $13$ TeV LHC is significantly smaller. Therefore, an $e^{+} e^{-}$ collider with large center of mass energy is more suitable to probe the high mass range.  For such heavy mass, the produced $W$s are boosted and their subsequent decay products will be collimated, resulting in  fat-jets. A number of SM processes, including  $4j$, $W^{\pm} 3j$, $W^{\pm} W^{\pm} 2j$ can mimic the signal. To reduce backgrounds, we carry out a jet-substructure analysis with $W$-tagging. We find that for the 800-1120 GeV mass range, a minimum of two tagged jets can effectively reduce the {total} backgrounds to a level of 
$\sigma \sim \mathcal{O}(0.1)$ fb, whereas the signal cross-section is {$\sigma \sim  \mathcal{O}(0.3-0.7)$ } fb. For higher masses, three tagged jets are needed. A doubly-charged Higgs boson with mass between 800-1120 GeV can be discovered with $\mathcal{L} \lsim 95 \,  \rm{fb}^{-1}$  of data. For even higher masses, such as $M_{H^{\pm \pm}} \sim 1400$ GeV, a discovery will require much higher integrated luminosities. 

Thus, a future high-energy $e^{+}e^{-}$ collider can provide an outstanding opportunity to probe weakly-coupled heavy particles, which are beyond the reach of the LHC.

\acknowledgments

MM acknowledges the support of the DST-INSPIRE research grant  IFA14-PH-99, and the cluster facility of Institute of Physics (IOP), Bhubaneswar, India. MM thanks the workshop 'Physics at CLIC', held during July 17th-18th at  CERN, Geneva, where part of the work has been carried out.   SS acknowledges the funding assistance provided by  Royal Society International Exchange program and the hospitality of the Institute for Particle Physics Phenomenology (IPPP) at Durham University where part of the work has been carried out.  
\bibliographystyle{JHEP}
\bibliography{t2seesaw.bib}

\end{document}